\documentstyle[aps,prl,multicol,amssymb,epsf]{revtex}

\title{The Halting Problem for Quantum Computers}

\author{N. Linden$^1$ and  S. Popescu$^{1,2}$}

\address{
$^1$Isaac Newton Institute for Mathematical
Sciences, Cambridge, CB3 0EH, UK\\%
$^2$BRIMS, Hewlett-Packard Laboratories, Stoke
Gifford, Bristol BS12 6QZ, UK\\
}

\date{26 June 98}
\begin{document}
\draft
\maketitle
\begin{abstract}
We argue that the halting problem for quantum computers  which was first raised 
by Myers, is by no means solved, as has been claimed recently.  We explicitly 
demonstrate the difficulties that arise in  a quantum computer when different 
branches of the computation halt at different, unknown, times.  
\end{abstract}


\begin{multicols}{2}


\newcommand\mathC{\mkern1mu\raise2.2pt\hbox{$\scriptscriptstyle|$}
                {\mkern-7mu\rm C}}		       
\newcommand{\mathR}{{\rm I\! R}}                

In \cite{Myers} Myers drew attention to the fact that there may be a problem 
if different branches of a quantum computation take different numbers of steps 
to complete their calculation.  In  a subsequent paper \cite{Ozawa}, Ozawa 
claimed, in effect, to have solved this problem.  We wish to reopen the issue.  
The reasons are two-fold.  Firstly we will show that the standard halting 
scheme 
for Turing machines which was also used 
in 
\cite{Ozawa} does not apply to any useful computers; the scheme is  unitary 
only for computers which do not halt.  Secondly, and more importantly, we will 
argue that the specific way the problem was framed, namely whether monitoring 
the halting does or does not spoil the computation, is not the important 
issue.
Indeed,  one can certainly build a quantum computer for which monitoring the 
halting does not spoil the computation by building a computer which is 
effectively classical. The key issue is whether the computer allows useful 
interference.

Later in this letter we will describe why the standard framework for quantum 
Turing 
machines 
as used in \cite{Ozawa} only applies 
to computers which do not halt.  First we set out, in general, what we would 
like a quantum computer to do.  

Any quantum 
algorithm relies on the fact that if an arbitrary input state $|i>$ evolves to 
the final 
state $|\psi_i>$ then the superposition $\sum_i a_i |i>$ evolves as
\begin{eqnarray}
\sum_i a_i |i> \mapsto \sum_i a_i |\psi_i>.
\end{eqnarray}
The states $|\psi_i>$ are, of course, not known beforehand; they arise at the 
end of the computation in which the computer  computes them, step by step, 
according to the program.   The problem is that for different inputs $|i>$, 
the number of steps required may not all be the same.  A situation might thus 
arise in which, say, the superposition 
\begin{eqnarray}
|1> + |2> 
\end{eqnarray}
 reaches
\begin{eqnarray}
|\psi_1> + |\tilde\psi_2> 
\end{eqnarray}
at a certain point during the computation, where $|\psi_1>$ is the final state
 obtained from $ |1>$,
but $|\tilde\psi_2>$ is not the final state which will be obtained from $|2>$, 
but just some intermediate result.  

What can one do?  An obvious suggestion is simply to wait until the 
computation in the second branch has also finished.  The problem is that 
unitarity of quantum evolution prevents  the state $|\psi_1>$ from remaining 
unchanged. To see this, let us denote by $U$ the time evolution operator 
corresponding to one step of computation.  We take $U$ to be time-independent, 
that is, we include all the relevant degrees of freedom as part of our 
computer. Otherwise we would have to take in to account the interactions of 
the computer with external degrees of freedom.  Let us suppose that 
$|\tilde\psi_1>$ is the state of the first branch a step before first reaching 
the final result. i.e.
\begin{eqnarray}
U|\tilde\psi_1> = |\psi_1>.\label{UPsiTilde}
\end{eqnarray}
Then it is impossible to also have
\begin{eqnarray}
U|\psi_1> = |\psi_1>.\label{HaltedPsi}
\end{eqnarray}
Indeed in order for both (\ref{UPsiTilde}) and (\ref{HaltedPsi}) to be true 
$|\tilde\psi_1>$ must be equal to $|\psi_1>$, but we have assumed that they are 
different by construction.

In order to allow the result of a computation to remain unchanged once a 
computation is finished, it is necessary to add some other degrees of freedom 
(i.e. an ancilla) to the computer which continue to evolve and thus preserve 
unitarity.
Following Deutsch \cite{Deutsch}, it is also customary to introduce a halt 
qubit 
which lies in a two-dimensional Hilbert space spanned by $|0>$ and $|1>$,
where $|0>$ means that the computation is still continuing and  $|1>$ means 
``halted''.   
We thus take the complete Hilbert space of a quantum computer to be spanned by 
vectors of the form
\begin{eqnarray}
|\psi>_C|H>_H |A>_A;
\end{eqnarray}
where the subscripts $C,\ H$ and $A$ refer to the computational states, the 
halt 
qubit and the ancilla respectively.

The introduction of an ancilla and the halt bit allows one to solve the problem 
of the halting of a given branch.  We 
require two conditions to be fulfilled for this branch.  Firstly, we require 
the computation to 
be able to stop. That is we require the computational state to change until it 
reaches its final state.  At that moment the halt qubit should change from 
$|0>$ 
to $|1>$.  The second requirement is that, once the computation has halted 
(i.e. 
once the halt qubit is in state $|1>$), neither the halt qubit nor the 
computational state should change further.  All that is allowed to change is 
the 
state of the ancilla.

These two requirements are written as follows:
\begin{eqnarray}
U|\tilde\chi_1>_{C,A}|0>_H=|\psi_1>_C|1>_H|a_0>_A\label{firsthalted}
\end{eqnarray}
and
\begin{eqnarray}
U|\psi_1>_C|1>_H|a_k>_A=|\psi_1>_C|1>_H|a_{k+1}>_A\label{haltedk}
\end{eqnarray}
where 
$|\tilde\chi_1>_{C,A}$  is the, possibly entangled, state of the computer and 
ancilla a step before first reaching 
the final result $|\psi_1>_C$, and $|a_k>_A$ are a set of states of the 
ancilla.
It is straightforward to see that unitarity requires 
\begin{eqnarray}
{}_A<a_k|a_{k^\prime}>_A=0\quad \forall k \neq k^\prime.
\end{eqnarray}

Thus once the computation has halted the ancilla starts evolving through a 
sequence of orthogonal states.  In effect, the ancilla contains a record of the 
time since the computation halted.  As we will show, this is at the core of the 
halting problem.  Roughly speaking, the computational states in two branches 
which halt at different, unknown, times do not interfere because they are 
entangled with this record of the halting time.

Within this general framework we can now discuss the different situations which 
might occur.  The simplest case is if all branches of the calculation halt at 
the same time.  In this case one can arrange things to have the desired 
interference of the computational states.  One way to do so is as follows. 
Take the state of the ancilla to be $|a_0>_A$, say, for all states at the 
beginning 
of the  computation.  The ancilla remains in this state until the point at 
which 
all the 
branches simultaneously halt.    We arrange exactly the same evolution of the 
ancilla for 
all
branches (i.e. the associated state of the ancilla remains 
$|a_0>_A$ until the halt qubit changes, then the ancilla starts to evolve in 
the  
sequence $|a_0>_A,\ |a_1>_A,\ |a_2>_A$ etc.).  We have thus arranged that there 
can be the required 
interference between branches. Thus in the situation where all branches halt at 
the same time, there is no difficulty.  It may also be noted that one can 
monitor the computer without spoiling the computation.

We now consider the  crucial case in which  different branches halt at 
different 
times, 
but we do not know in advance how long each branch takes.
Consider two branches which halt at different times and assume that the first 
branch halts first.  It then evolves in the following way:
\begin{eqnarray}
& &|\psi_1>_C|1>_H|a_0>_A\nonumber\\
& &\mapsto |\psi_1>_C|1>_H|a_{1}>_A\nonumber\\
& &\mapsto |\psi_1>_C|1>_H|a_{2}>_A\quad\ldots
\end{eqnarray}
The second branch halts at some later time and then evolves as
\begin{eqnarray}
& &|\psi_2>_C|1>_H|b_0>_A\nonumber\\
& &\mapsto |\psi_2>_C|1>_H|b_{1}>_A\nonumber\\
& &\mapsto |\psi_2>_C|1>_H|b_{2}>_A\quad\ldots
\end{eqnarray}
Here the $|b_k>$ are some other orthogonal set of states of the 
ancilla which need not be related to the $|a_k>$ (we note that 
(\ref{firsthalted}) and 
(\ref{haltedk}) do not require that the set of states through which the ancilla 
evolves 
after halting be the same for different branches).

The simplest possibility is to arrange that the sequence of states of the 
ancilla 
is the 
 same for all branches (i.e. namely
the  state of the ancilla remains 
$|a_0>_A$ for all states until the halt qubit changes, then the ancilla starts 
to evolve in the  
sequence $|a_0>_A,\ |a_1>_A,\ |a_2>_A$ etc.).  
However, although the ancilla evolves through exactly the same set of states 
for 
all 
branches, branches which halt at different times  are not synchronized.
Consequently two computational branches which halt at different times
first decohere because they are entangled with two different states of the halt 
bit
but even after both branches have halted, they still decohere due to 
entanglement 
to 
orthogonal states of the ancilla.

This situation has the property that 
monitoring the computation does not affect it and is thus an example of how it 
can be arranged that monitoring the calculation does not spoil it.  
However the reason that this is the  case is that, in fact, 
there is no interference at all between branches which halt at 
different times even in the absence of monitoring. Thus as far as branches 
which 
halt at different times are 
concerned, the computation is effectively classical.

Alternatively it might be possible, for example, to choose the $|b_k>$ to be 
the 
same 
set as
the $|a_k>$, but in some different order.  At certain times in the future
there might be reinterference of different branches; however these times are 
unknown.  We also note that in this situation the monitoring the halting bit 
does 
affect the results of the computation as it prevents reinterference of the 
computational bits.  However monitoring is not the important issue; the issue 
is 
that although (in the absence of monitoring) there might be reinterference, 
since 
one does not know when it occurs, it is not useful for computational purposes.

Other choices for the $|b_k>$ can be made but not so as to arrange useful 
interference.

We now turn to a discussion of the standard halting scheme for quantum Turing 
machines as 
in  \cite{Ozawa}.  We will argue that this  
halting scheme  is not consistent with unitarity except in the trivial 
case in which the computer never halts. (As will become clear, the problems 
with this halting scheme are independent of the issue of whether 
the branches halt at the same or different times.)

Following the  discussion in \cite{Ozawa} we write the state of a
 quantum Turing machine. 
 in terms of a  basis
\begin{equation}
|C> = |q_C>|h_C>|T_C>|H>.
\end{equation} 
$|q_C>$ is the internal state of the head, assumed, by definition,
 to lie in a 
finite-dimensional Hilbert space and $|h_C>$ is the position of the head.  
$|T_C>$ is the state of the tape; the tape is built out of cells, each cell 
carrying an identical finite-dimensional Hilbert space.  In addition the 
system has a halt qubit $|H>$
which lies in a two-dimensional Hilbert space spanned by $|0>$ and $|1>$,
where $|1>$ means ``halted''.  The evolution of the computer occurs in steps; 
each step being described by the same unitary operator $U$.  This unitary 
operator is such that the internal state of the head, the state of the tape 
cell at the location of the head, the state of the halting bit and the 
position of the head are updated according to the current state of the head, 
the state of the qubit at the current position of the head and the current 
state of the halting qubit.

The key equation describing the halting scheme is equation 
(6) of 
\cite{Ozawa}:
\begin{eqnarray}
& & U|q_C>|h_C>|T_C>|1>\nonumber\\
& &\quad = \sum_{q,d}c_{q,d}|q>|h_C+d>|T_C>|1>,\label{Ozawa6}
\end{eqnarray}
where the quantity $d$ may have values $+1$ or $-1$ denoting whether the head 
has moved to the right or left, and $c_{q,d}$ are constants.  According to 
(\ref{Ozawa6}) once the halt qubit is set to
$|1>$, the proposed quantum Turing machine no longer changes the halt qubit or 
the tape string.

The above halting scheme seems very natural, however it contains subtle but 
very serious problem. We will show that (\ref{Ozawa6}) implies that the 
unitary evolution operator 
$U$
is essentially trivial, namely that (\ref{Ozawa6}) cannot be satisfied by any 
$U$ which allows the halt bit of any state to change from $|0>$ to $|1>$. 
i.e. the halting scheme is valid only for a computer which never halts.

In order to prove our result, let us first consider the following set of 
states 
in which the halt qubit is $|1>$.
\begin{eqnarray}
|q_j>|n>|{\bf  \hat T},T_n=\xi,T_{n+2}=\xi>|1>,\label{halted}
\end{eqnarray}
where the states $|q_j>$, $j=1...M$ are an orthonormal basis for the internal 
states, and where $|n>$, $-\infty<n<\infty$ are states labeled by the integer 
$n$ 
specifying the position of the 
head. 
$|{\bf  \hat T},T_n=\xi,T_{n+2}=\xi>$ is the state of the tape; ${\bf  \hat 
T}$ 
labels 
the states of the cells at all positions on the tape except those explicitly 
exhibited, 
in this 
case the cells at positions $n$ and $n+2$ where the states are $\xi$.

The most general evolution of (\ref{halted}) under (\ref{Ozawa6}) is
\begin{eqnarray}
& & U|q_j>|n>|{\bf  \hat T},T_n=\xi,T_{n+2}=\xi>|1>\nonumber\\
& &\quad =
(|Q^+_j>|n+1> + |Q^-_j>|n-1>)\nonumber\\
& &\qquad\times|{\bf  \hat T},T_n=\xi,T_{n+2}=\xi>|1>,\label{Uhalted}
\end{eqnarray}
where we have not assumed that the states $|Q^+_j>$ and $|Q^-_j>$ are  
necessarily normalized or orthogonal.

For two different values of $j$, states of the form (\ref{halted}) are 
orthogonal so the states to which they evolve must also be orthogonal; also 
the 
norm of a given state must not change under evolution.  Thus one derives that
\begin{equation}
<Q^+_j|Q^+_k> + <Q^-_j|Q^-_k> = \delta_{jk}.\label{condition1}
\end{equation}
Let us now consider the following state
\begin{equation}
|q_k>|n+2>|{\bf  \hat T},T_n=\xi,T_{n+2}=\xi>|1>;\label{haltedplus2}
\end{equation} 
this evolves to
\begin{eqnarray}
& &(|Q^+_k>|n+3> + |Q^-_k>|n+1>)\nonumber\\
& &\quad\times |{\bf  T},T_n=\xi,T_{n+2}=\xi>|1>.\label{Uhaltedplus2}
\end{eqnarray}
Note that $|Q^+_k>$ and $|Q^-_k>$ are the same set of states as in appear in 
(\ref{Uhalted}) 
since the evolved 
state can only depend on the internal state of the head and the tape at the
position of the head.

Now (\ref{halted}) and (\ref{haltedplus2}) are orthogonal for all $j$ and $k$;
thus
\begin{eqnarray}
<Q^-_j|Q^+_k>=0\quad \forall j,k.\label{condition2}
\end{eqnarray}
Now consider the $M$ states 
\begin{eqnarray}
|E_k> := |Q^+_k> + |Q^-_k>\label{Ek}
\end{eqnarray}
The conditions (\ref{condition1}) and (\ref{condition2}) imply that
\begin{eqnarray}
<E_j|E_k>=\delta_{jk}. 
\end{eqnarray}
Since $j$ and $k$ run from $1$ to $M$ (the dimension of the space) the $|E_k>$ 
form an orthonormal basis for the internal states of the 
head.

Let us now consider the action of $U$ on a state which has the halt qubit in 
the state $|0>$.   For the halt scheme to be non-trivial there must exist at 
least one internal state of the head $|q_0>$, say, and a  tape-cell state 
$\eta$, such that if internal state of the head is $|q_0>$ and the state of 
cell at the current position of the head is $\eta$ then the  state with halt 
qubit set to $|0>$ evolves to a state which has a least one component  with 
the halt qubit set to $|1>$.  

Let us now consider the state 
\begin{equation}
|q_0>|n>|{\bf  \hat T}, T_n = \eta,T_{n+2} = \psi>|0>,\label{nonhalted}
\end{equation} 
where we have labeled the state at $n+2$ for later convenience.
The most general way in which (\ref{nonhalted}) can evolve is
\begin{eqnarray}
& &U|q_0>|n>|{\bf  \hat T}, T_n = \eta,T_{n+2} = \psi>|0>\nonumber\\
& &\quad =\sum_\mu |\Phi^-_\mu>|n-1>|{\bf  \hat T}, T_n = e_\mu,T_{n+2} = 
\psi>|1>\nonumber\\
& &\qquad + \sum_\mu |\Phi^+_\mu>|n+1>|{\bf  \hat T}, T_n = e_\mu,T_{n+2} = 
\psi>|1>
\nonumber\\
& &\qquad + |\Psi>|0>.\label{Unonhalted}
\end{eqnarray}
The states $| e_\mu>$ are an orthonormal basis for the states of the tape at 
the 
position $n$ and $|\Phi^-_\mu>$ are generic (non-normalized, and not 
necessarily mutually orthogonal) internal states so 
that $\sum_\mu |\Phi^-_\mu>|t_n = e_\mu>$ is the most general entangled state 
of the tape cell at $n$ and the internal states of the head.  $|\Psi>$ is some 
 state of the position of head, the tape and the internal state of the head 
which need not be specified. Note 
that the output states of the head, $|\Phi^\pm_\mu>$ are independent of the 
states of the tape except at $n$, hence in particular they are independent of
$\psi$.

Now consider a halted state with the state of the tape chosen to be
$| e_\nu>$,  at the positions $n$ and $n+2$, i.e.
\begin{eqnarray}
|q_j>|n>|{\bf  \hat T},T_n=e_\nu,T_{n+2}=e_\nu>|1>.
\end{eqnarray}
This state is orthogonal to the state (\ref{nonhalted}) for any $\psi$ and for 
$\psi = e_\nu$ in particular, so that the fact that the states must also be 
orthogonal after evolution by $U$ 
requires that
\begin{equation}
<Q^+_j|\Phi^+_\nu> + <Q^-_j|\Phi^-_\nu> = 0 \quad\forall 
\mu,j.\label{condition3}
\end{equation}
Also the fact that (\ref{nonhalted}), with $\psi$ chosen to be $e_\nu$,
 is orthogonal to 
\begin{equation}
|q_j>|n+2>|{\bf  \hat T},T_n=e_\nu,T_{n+2}=e_\nu>|1>;
\end{equation} 
shows that
\begin{equation}
<Q^-_j|\Phi^+_\nu> = 0 \quad\forall \nu,j.\label{condition4}
\end{equation}
Similarly, by considering a suitably chosen halted state at position $n-2$,
we may show that
\begin{equation}
<Q^+_j|\Phi^-_\nu> = 0 \quad\forall \nu,j.\label{condition5}
\end{equation}
Now  (\ref{condition3}), (\ref{condition4}) and (\ref{condition5})
imply that
\begin{equation}
|\Phi^+_\nu> + |\Phi^-_\nu> 
\end{equation}
is orthogonal to all basis vectors $|E_k>$  and so 
\begin{equation}
|\Phi^+_\nu> + |\Phi^-_\nu> = 0.
\end{equation}
Thus substituting $|\Phi^+_\nu> = - |\Phi^-_\nu>$ in (\ref{condition4})
we find 
\begin{equation}
<Q^-_j|\Phi^-_\nu> = 0 \quad\forall \nu,j.\label{condition6}
\end{equation}
and so from (\ref{Ek}),(\ref{condition5}) and (\ref{condition6})
$|\Phi^-_\nu>$ is orthogonal to $|E_k>$ for all $k$ and hence 
we find 
\begin{equation}
|\Phi^-_\nu> =|\Phi^+_\nu> = 0 \quad\forall \nu\label{Phizero}.
\end{equation}
Thus considering (\ref{Unonhalted}) and (\ref{Phizero}) we reach our 
conclusion that the requirement that $U$ be unitary prevents
any state from evolving from un-halted to halted.  Thus the halting scheme and 
therefore the results in \cite{Ozawa} only apply 
to computers which do not halt.

It  might be wondered why this particular halting scheme does not work, 
although 
it seems to be almost identical to the one presented earlier in this letter.  
The internal states of the machine and the position of the head appear to play 
the role of the ancilla in our model.  However these degrees of freedom play an 
active role during computation and have highly constrained dynamics and it is 
this which creates the problem.

In conclusion, we have shown that the halting problem for quantum computers is 
by 
no means solved as has been claimed.  In particular we have shown that the 
standard halting scheme for quantum Turing machines, as used in \cite{Ozawa} is 
not consisent with unitarity. We have also given a general discussion of 
halting in quantum computers illustrating the problems which arise when 
different branches of the computation halt at different times.

We should point out that our discussion of the 
general problem of halting is not exhaustive.  For example we have considered a 
particular model of  quantum computation in which when a branch halts, it halts 
with certainty in a single step.  We have shown, that in this model, if the 
halting time of different branches of the computer are different and unknown,
then useful interference is not possible. We anticipate that similar  problems 
will arise in any model in which branches halt at unknown times.  One possible 
resolution, as has been discussed by Bernstein and Vazirani \cite{BV}, is to 
restructure algorithms so as to ensure simultaneous halting.  We do not know, 
however, whether this can be done for all computational problems.

\bigskip
\noindent{\large\bf Acknowledgments}
We are grateful to Yu Shi and Richard Jozsa for very helpful discussions.

\end{multicols}

\end{document}